
\input phyzzx
%
\catcode`\@=11
\paperfootline={\hss\iffrontpage\else\ifp@genum\tenrm
 -- \folio\ --\hss\fi\fi}
\def\titlestyle#1{\par\begingroup \titleparagraphs
 \iftwelv@\fourteenpoint\fourteenbf\else\twelvepoint\twelvebf\fi
 \noindent #1\par\endgroup }
\def\GENITEM#1;#2{\par \hangafter=0 \hangindent=#1
 \Textindent{#2}\ignorespaces}
\def\address#1{\par\kern 5pt\titlestyle{\twelvepoint\sl #1}}
\def\abstract{\par\dimen@=\prevdepth \hrule height\z@ \prevdepth=\dimen@
 \vskip\frontpageskip\centerline{\fourteencp Abstract}\vskip\headskip }
\newif\ifYUKAWA  \YUKAWAtrue
\font\elevenmib   =cmmib10 scaled\magstephalf   \skewchar\elevenmib='177
\def\YUKAWAmark{\hbox{\elevenmib
 Yukawa\hskip0.05cm Institute\hskip0.05cm Kyoto \hfill}}
\def\titlepage{\FRONTPAGE\papers\ifPhysRev\PH@SR@V\fi
 \ifYUKAWA\null\vskip-1.70cm\YUKAWAmark\vskip0.6cm\fi
 \ifp@bblock\p@bblock \else\hrule height\z@ \rel@x \fi }

\def\schapter#1{\par \penalty-300 \vskip\chapterskip
 \spacecheck\chapterminspace
 \chapterreset \titlestyle{\ifcn@@\S\ \chapterlabel.~\fi #1}
 \nobreak\vskip\headskip \penalty 30000
 {\pr@tect\wlog{\string\chapter\space \chapterlabel}} }

\def\ssection#1{\par \ifnum\lastpenalty=30000\else
 \penalty-200\vskip\sectionskip \spacecheck\sectionminspace\fi
 \gl@bal\advance\sectionnumber by 1
 {\pr@tect
 \xdef\sectionlabel{\ifcn@@ \chapterlabel.\fi
 \the\sectionstyle{\the\sectionnumber}}%
 \wlog{\string\section\space \sectionlabel}}%
 \noindent {\S \caps\thinspace\sectionlabel.~~#1}\par
 \nobreak\vskip\headskip \penalty 30000 }


\papers

\def\lkakko{\vbox{\vskip0.065cm\hbox{(}\vskip-0.065cm}}
\def\rkakko{\vbox{\vskip0.065cm\hbox{)}\vskip-0.065cm}}
\def\YUKAWAHALL{\hbox to \hsize
 {\hfil \lkakko\twelvebf YUKAWA HALL\rkakko\hfil}}





\def\addeqno{\ifnum\equanumber<0 \global\advance\equanumber by -1
 \else \global\advance\equanumber by 1\fi}


\mathchardef\Lag="724C
\def\sqr#1#2{{\vcenter{\hrule height.#2pt
 \hbox{\vrule width.#2pt height#1pt \kern#1pt\vrule width.#2pt}
 \hrule height.#2pt}}}


\def\cref#1{\rlap,\attach{#1)}}
\def\ref#1{\attach{#1)}}



\newdimen\ex@
\ex@.2326ex
\def\boxed#1{\setbox\z@\hbox{$\displaystyle{#1}$}\hbox{\lower.4\ex@
 \hbox{\lower3\ex@\hbox{\lower\dp\z@\hbox{\vbox{\hrule height.4\ex@
 \hbox{\vrule width.4\ex@\hskip3\ex@\vbox{\vskip3\ex@\box\z@\vskip3\ex@}%
 \hskip3\ex@\vrule width.4\ex@}\hrule height.4\ex@}}}}}}
\def\txtboxed#1{\setbox\z@\hbox{{#1}}\hbox{\lower.4\ex@
 \hbox{\lower3\ex@\hbox{\lower\dp\z@\hbox{\vbox{\hrule height.4\ex@
 \hbox{\vrule width.4\ex@\hskip3\ex@\vbox{\vskip3\ex@\box\z@\vskip3\ex@}%
 \hskip3\ex@\vrule width.4\ex@}\hrule height.4\ex@}}}}}}
\newdimen\exx@
\exx@.1ex
\def\thinboxed#1{\setbox\z@\hbox{$\displaystyle{#1}$}\hbox{\lower.4\exx@
 \hbox{\lower3\exx@\hbox{\lower\dp\z@\hbox{\vbox{\hrule height.4\exx@
 \hbox{\vrule width.4\exx@\hskip3\exx@%
 \vbox{\vskip3\ex@\box\z@\vskip3\exx@}%
 \hskip3\exx@\vrule width.4\exx@}\hrule height.4\exx@}}}}}}

\chardef\fontD="1A

\catcode`@=12
\rightline{KOBE-TH-95-05}
\vskip -2mm
\rightline{KEK-TH-448}
\vskip9mm
\pubnum={YITP/K-1101}
\date={July 1995}

\titlepage
\title{Non-decoupling Effects of Heavy Particles
in Triple Gauge Boson Vertices}

\author{Takeo Inami\cref{a} C.S. Lim\cref{b}
 B. Takeuchi\cref{b}$^,$\footnote{\dagger}
{Present address: Shikata-cho 849, Kakogawa, Hyogo 675-03, Japan}
M. Tanabashi\ref{c}}

\vskip3mm
\line{\sl \hbox to 5mm{}
{\rm a} ~Yukawa Institute for Theoretical Physics,
 Kyoto University,~Kyoto 606,~Japan \hfill}

\line{\sl \hbox to 5mm{}
{\rm b} ~Department of Physics, Kobe University, Kobe 657,~Japan \hfill}

\line{\sl \hbox to 5mm{}
{\rm c} ~~National Laboratory for High Energy Physics, Oho 1-1,
Tsukuba 305,~Japan \hfill}

\abstract{
Non-decoupling effects of heavy particles
present in beyond-the-standard models are
studied for the triple gauge boson vertices
$\gamma W^+W^-$ and $Z^0W^+W^-$.
We show from a general argument that the non-decoupling
effects are described by four independent
parameters, in comparison with the three parameters
$S$, $T$ and $U$ in
the oblique corrections.
These four parameters
of the effective triple gauge
boson vertices
are computed in two beyond-the-standard
models.  We also study the relation of the four
parameters to the $S$, $T$, $U$ parameters,
relying on an operator analysis.
}

\endpage

\null
\vskip-4.5mm
\centerline{\fourteenbf 1. Introduction}
\vskip1mm
Any unified model of electro-weak (and strong)
interactions beyond the standard model is
characterized by the existence of heavy particles
of masses $M \gg M_W$,
$M_W$ being the weak-boson mass.
Since such heavy particles are not likely
to be discovered in the immediate future,
it is of importance to ask how their effects may
be detected in low energy
($ E \lsim M_W$) processes of
light ($m \lsim M_W$) particles,
such as those measured at LEP and SLC
experiments, through radiative
corrections.  In this paper we will restrict
ourselves to those classes of
beyond-the-standard models
in which heavy particles manifest
themselves in low energy processes only
through loop effects.

It is useful to divide the loop
effects of heavy particles (the new
physics contributions) into two types:
i) those which virtually decouple in
the limit $M \rightarrow \infty$
and ii) those which do not
decouple in the same limit.

In the type i) the heavy mass $M$ is
dominated by a new large mass scale $M_S$,
{}~$M^2 = M^2_S + O(M^2_W)$.
Note that the term giving $M_S$ is
$SU(2)_L \times U(1) $
singlet, and hence the heavy particle
contributions to low-energy processes are
suppressed by $1/{M^2_S}$
[1].
On the other hand, in the type ii) $M$
has its origin in the $SU(2)_L \times U(1)$
breaking due to the VEV of the Higgs
$\phi$, and large $M$ means a large coupling
constant.  The latter factor appearing in
the numerator of amplitudes cancels the
suppression factor $1/{M^2}$,
leading to non-decoupling effects of heavy
particles.

We will be concerned with non-decoupling
effects of the type ii), which will
provide us with some useful constraints on the
properties of new physics which might
lie beyond the standard model.
As far as light fermion processes are
concerned, we only have to deal with
the non-decoupling contributions
to the gauge boson two-point functions
(oblique correction [2]).
These corrections are summarized in terms
of three parameters $S$, $T$ and $U$[3-6].
A few implications on beyond-the-standard models
have been obtained recently.  For instance,
realistic technicolor models have
been shown to contradict the observed value of the
$S$ [3,6].
We have derived a bound on the number of possible heavy
extra generations of fermions by a combined use
of recent data on the
$S$ and $T$ [7].

The next question to ask is whether
similar non-decoupling contributions of heavy particles are
present in higher $n$-point functions of gauge bosons.
The answer is known.  Namely, (at
one-loop level) for $n>4$ the
coefficient of the relevant operator $O_i$ is
suppressed by $1/{M^{n-4}}$
or more strongly since $O_i$ has dimension
$d_i \geq n$.
Hence $n = 2,~3,~4$ are the only possibility to
have non-decoupling effects.
In this paper we wish to study the
non-decoupling effects in the triple gauge
boson (TGB) vertices ($n=3$).
These vertices can be
measured in the $W^+W^-$ production at
LEP200.

The purpose of the present paper is to
identify the minimum set of parameters
which describe the non-decoupling effects of heavy
particles in the TGB vertices, just as the
$S$, $T$ and $U$ parameters do in the oblique
corrections.  We will show
by a general argument that the
non-decoupling effects in the TGB vertices are
summarized in terms of four
parameters.  This conclusion from a general
analysis will be confirmed by the
evaluation of the four parameters at
one-loop level in two examples
of beyond-the-standard model, (a) an
extra fermion generation and (b) technihadrons.

We may easily understand why the non-decoupling
effects in the TGB vertices are inevitable from
a simple operator analysis.  In a theory with
spontaneous gauge symmetry breaking,
quantum corrections due to
heavy loops can be described in terms of gauge
invariant effective operators consisting
of the gauge fields and the Higgs field
$\phi$.  Gauge non-invariant
effective operators in the
broken phase arise when $\phi$ is replaced by the
sum of its VEV $v$ and the shifted field, as
exemplified by the operator of
dimension 6 [8,9],

$$
(\phi^{\dagger} \sigma^a \phi)W^a_{\mu \nu}B^{\mu \nu}
= v^2W^3_{\mu \nu}B^{\mu \nu}~~,\eqno{(1)}
$$
\noindent
where $W^3_{\mu \nu} = \partial_\mu W^3_\nu - \partial_\nu W^3_\mu
-igW^+_\mu W^-_\nu$.
The $r.h.s.$ of eq.(1) is an expression in the broken
symmetry phase.
The first term of $W^3_{\mu\nu}$
contributes to the $S$ parameter, while the
second induces a TGB coupling,
$W^+_\mu W^-_\nu B^{\mu\nu}$.

The question how the parameters describing the
non-decoupling effects in the TGB
couplings are related to the parameters
$S$, $T$ $U$ depends on whether
the new physics contributions in
question are of the type i) or ii) mentioned
above.  This question will be answered later.

\vskip11mm
\centerline{\fourteenbf 2. Triple gauge boson vertices and four parameters}

In the present paper we consider two kinds of triple gauge boson (TGB) vertices
$VW^+W^-$ where $V$ denotes photon $\gamma$ or $Z^0$.
These couplings can be measured by observing
the $W^{\pm}$ pair production in $e^+e^-$
collisions,
$$
e^-+e^+ \rightarrow W^- + W^+~~.\eqno{(2)}
$$

Both neutrino exchange in the $t$-channel and $V$
exchange in the $s$-channel contribute
to this process (Fig. 1).
In the $t$-channel process the heavy
particle effects are confined to
the loop correction on the external $W^{\pm}$
legs.  The results can be found in the
literature [10], and we will not discuss them
further.  The $V$ exchange diagram consists
of three factors, the $V$ propagator $\Pi$,
the $VW^+W^-$ vertices
$\Gamma^V$ and the external leg
corrections $ \Pi^{'}_{11} $ of
$W^{\pm}$ (Fig.1).
The heavy particle effects in the
$V$ propagator have already been taken
into account as the oblique corrections.
It suffices to consider the
$VW^+W^-$ vertices $\Gamma^V$.

We define the kinematics of the
$VW^+W^-$
vertex as
$$
V(p,~\epsilon_1) \rightarrow W^-(q,~\epsilon_2)
+ W^+(\bar{q},~\epsilon_3)~, \eqno{(3)}
$$
\noindent
where $p,~q,~\bar{q}$
are the momenta of $V,~ W^-,~ W^+$
respectively and $ \epsilon_1,~\epsilon_2,~\epsilon_3$
their polarization vectors.
The produced $W^{\pm}$
are on-shell and we impose
$$
q^2 = \bar{q}^2 = M^2_W~,~~~ q\cdot\epsilon_2
=\bar{q}\cdot\epsilon_3=0~~.\eqno{(4)}
$$
We are interested in the low-energy process, which means
that
$$
p^2 = {(q+ \bar{q})}^2~,~~~~q^2=
\bar{q}^2 \ll M^2~~.\eqno{(5)}
$$
The process (2) in the high energy limit, $p^2 \gg M^2$
has been studied in [10].
In many of the models with heavy particles, at
one-loop level, the $VW^+W^-$ vertex preserves
$CP$ invariance, which we assume
in this article.

Let $g_{WWV}$ $\Gamma^V_{\mu\alpha\beta}$
be the effective $VW^+W^-$ vertex, where
$g_{WW\gamma}=-e,~~g_{WWZ}=-e\cot \theta_W$.
Following Hagiwara, Hikasa, Peccei and
Zeppenfeld [11], we
decompose the effective vertex into
pieces with different Lorentz structures,
$$
\eqalign{
\Gamma^V_{\mu \alpha \beta} =
& f^V_1{(q- \bar{q})}_\mu g_{\alpha \beta} +
f^V_2{(q- \bar{q})}_\mu p_\alpha p_\beta +
f^V_3{(g_{\mu \beta}p_\alpha - g_{\mu \alpha} p_\beta)} \cr
& + if^V_5 \epsilon_{\mu \alpha \beta \rho}
{(q- \bar{q})}^\rho + ih^Vp_\mu \epsilon_{\alpha \beta \rho \sigma}
p^\rho{(q- {\bar{q}})}^\sigma~~ \cr
}\eqno{(6)}
$$

\noindent (we take the convention $\epsilon^{0123}=1$). We have imposed the
on-shell conditions for $W^{\pm}$ and $CP$ invariance.
The terms proportional to $p_\mu$ have been
ignored, since they give terms proportional
to external electron masses on using the
equation of motion.  The last term with
$h^V$ is redundant in this sense, but it is kept to
make unbroken $U(1)_{em}$ gauge invariance manifest
in the case of $V= \gamma$.

At the tree level $f^V_1$ and $f^V_3$ are the
only non-vanishing form factors,
$$
f^{\gamma}_1 = f^Z_1 =1~~,~
{}~~~f^{\gamma}_3 = f^Z_3 =2~~.
\eqno{(7)}
$$
This in turn means that the one-loop corrections
to $f^V_1$ and $f^V_3$ are apparently
ultraviolet-divergent.
When the external leg correction, or the wave-function
renormalization factor of $W^{\pm}$
$$
Z_W = 1 + g^2 \Pi^{'}_{11}~,
\eqno{(8)}
$$
is combined, the divergent form factors
are rendered finite.  Namely,
$$
\eqalign{
f^V_1 + g^2 \Pi^{'}_{11} & \equiv 1 + \bar{f}^V_1~,\cr
f^V_3 + 2g^2 \Pi^{'}_{11} & \equiv 2 + \bar{f}^V_3~, \cr
}
\eqno{(9)}
$$
where $\bar{f}^V_i$ are the finite
one-loop corrections to $f^V_i$.

This prescription to get finite form
factors is somewhat different from that
adopted in Ref.[10], where
the $W^3{\gamma}$ self-energy
$ \Pi^P_{3Q}$ is chosen, instead of
$\Pi^{'}_{11}$,
to get the finite form factors.
The ``charge universality" is manifest
in our choice of form factors, i.e.
$\bar{f}^{\gamma}_1=0$
as will be discussed below, while
$f^{\gamma}_1 + g^2 \Pi^P_{3Q} \neq 1$.

We are now ready to show how many
parameters are necessary to describe the
non-decoupling effects in the
$VW^+W^-$ vertices.
We follow the way of argument by
Altarelli and Barbieri concerning the
oblique corrections [5].
First we note that in the expansion of each form
factor $f^V_i({p^2})$
in the powers of
$p^2/M^2$,
only the lowest term survives, since
all form factors are
at most dimensionless.
Thus we are left with the parameters
$f^V_i(0)$
as the candidates,
which we will refer to simply as
$f^V_i$ hereafter.
The parameter $f^V_2$ actually decouples,
since it has mass dimension -2 and therefore is
suppressed by $1/{M^2}$.

Next we have to take account of the $U(1)_{em}$
gauge invariance.
This amounts to imposing the current conservation,
$p^{\mu}\Gamma^{\gamma}_{\mu \alpha \beta} = 0$.
Under the on-shell condition for $W^{\pm}$,
only $f^{\gamma}_5$
is subject to a non-trivial condition,
$$
f^{\gamma}_5 (p^2) = (-p^2) h^{\gamma}~~.
\eqno{(10)}
$$
Since $h^{\gamma}$ has no singularity at
$p^2=0$, we conclude that
$f^{\gamma}_5(p^2)$ is suppressed by
${p^2}/{M^2}$
and decouples.

We should also consider the condition which arises
as a result of ``charge universality"
of the photon coupling.  In QED the electric charge
defined in the Thomson limit $p^2=0$
has a universal meaning, i.e. irrespectively of
external particles;
it is fixed by $Z_3$, the wave-function
renormalization of the photon.  In the
$SU(2)_L \times U(1)$
theory, the non-Abelian nature is inessential
in this respect, and hence the electric charge
is fixed by the
oblique corrections alone; the vertex correction
to $f^{\gamma}_1$ and the external leg corrections
of $W^{\pm}$ should cancel
out in the ``effective Thomson limit"
$p^2 \ll M^2$, i.e.
$$
\bar{f}^{\gamma}_1 = f^{\gamma}_1(0) +
g^2 \Pi^{'}_{11}(0) - 1 = 0~~.
\eqno{(11)}
$$

To summarize, we have shown
that there remain four independent
parameters
$$
\bar{f}^Z_1~~, \bar{f}^{\gamma}_3~~,\bar{f}^Z_3~~, f^Z_5
\eqno{(12)}
$$
in order to describe the non-decoupling
loop corrections to the TGB couplings due
to heavy particles.

Experimental constraints on the
four parameters may be obtained from
precise measurements of
$e^-+e^+ \rightarrow W^-+W^+$
process.
The four parameters participate in the
$s$-channel matrix element.
As for the loop corrections to the $V$
propagator,
the oblique corrections, they can be taken
into account by replacing electric charge $e$,
Weinberg angle $s$, and $M_Z$ by their
corresponding star quantities
$e_{\ast},~s_{\ast},~M_{Z^\ast}$
at the tree level amplitude [2,3,10].
In a similar way, the sum of
remaining vertex correction and the
external leg correction of $W^{\pm}$
can be taken account of by
replacing $\Gamma^V_{\mu \alpha \beta}$
at the tree level by, say,
$\Gamma^{\ast V}_{\mu \alpha \beta}$, where
$$
\eqalign{
\Gamma^{\ast \gamma}_{\mu \alpha \beta}
&= {(q- \bar{q})}_{\mu} g_{\alpha \beta}
+ (2 + \bar{f}^{\gamma}_3)
(g_{\mu \beta} p_\alpha - g_{\mu \alpha}p_\beta), \cr
\Gamma^{\ast Z}_{\mu \alpha \beta}
&= (1 + \bar{f}^Z_1) {(q- \bar{q})}_{\mu}
g_{\alpha \beta} + (2 + \bar{f}^Z_3)
(g_{\mu \beta} p_{\alpha} - g_{\mu \alpha} p_{\beta}) \cr
&+ if^Z_5 \epsilon_{\mu \alpha \beta \rho} {(q- \bar{q})}^{\rho}~~. \cr
}
\eqno{(13)}
$$
Thus the loop-corrected
$s$-channel matrix element is given
as
$$
\eqalign{
M = &-ie^2_{\ast}Q (\bar{v}\gamma^{\mu}u)
{1 \over {p^2}} \Gamma^{\ast \gamma}_{\mu \alpha \beta}
{\epsilon^{\alpha}_2}(q) {\epsilon^{\beta}_3}(\bar{q}) \cr
&-ie^2_{\ast}
{ {(I_3 - s^2_{\ast}Q)} \over s^2_{\ast}}
(\bar{v}\gamma^{\mu}u)
{ 1 \over { p^2 - {M^2_{Z^{\ast}}}}}
\Gamma^{\ast Z}_{\mu \alpha \beta}
{\epsilon^{\alpha}_2}(q) {\epsilon^{\beta}_3}(\bar{q})~~~,
\cr
}
\eqno{(14)}
$$
where $u$ and $v$ are electron and positron
(Weyl) spinors, and
${\epsilon^{\alpha}_2}(q)$
and ${\epsilon^{\beta}_3}(\bar{q})$
$W^{\pm}$
polarization vectors.  In eq.(14),
$e^+$ and $e^-$ are assigned a
definite helicity, and
$I_3 = -{1 \over 2}$ for
$e_L$, $I_3 =  0$ for $e_R~ (Q =-1)$.

The four form factors get contributions
from ordinary particles of the
standard model as well.
These contributions of the standard
model [12] have to be subtracted from the
measured values of the form factors,
when one tries to derive
experimental constraints on the
non-decoupling effects of heavy particles.

\vskip11mm
\centerline{\fourteenbf 3. New physics contributions to the four parameters}

Here we evaluate the four non-decoupling parameters
by taking two examples of new physics contributions:
(a) an extra fermion generation and (b) technihadrons.

\vskip3mm
(a) {\it Extra fermion generation}

Consider an extra
generation of quarks and leptons ($Q$ and $L$):
$$
{U \choose D}_L~~,~~
U_R~~,~D_R~~;
{N \choose E}_L~~,~~
N_R~~,~~E_R~~.
\eqno{(15)}
$$
\noindent They are assumed to be a color triplet and a singlet
(the color factor $N_Q=3$ and $N_L=1$) and to have the hypercharge
$Y_Q=1/6$ and $Y_L=-1/2$ (for the $SU(2)_L$ doublet), respectively.
The neutral
lepton N is assumed to have a Dirac mass.  The fermions are
all assumed to be heavy,
$m^2_F \gg M^2_W$.
The one-loop graphs of the TGB
vertices are ultraviolet-divergent and have
to be regularized.
We have computed the parameters
$\bar{f}^Z_1$, $\bar{f}^{\gamma}_3$,
$\bar{f}^Z_3$
in the dimensional regularization with
anti-commuting $\gamma_5{ ( \{ \gamma_\mu,~\gamma_5 \} = 0 )}$,
which obviously respects gauge
symmetry.  As for $f^Z_5$,
the usual dimensional method gives
vanishing $f^Z_5$.
This is because of the identity
$Tr~{( \gamma_\mu \gamma_\alpha
 \gamma_\beta \gamma_\rho \gamma_5 )} = 0$,
the same reason that the chiral anomaly cannot
be obtained in this regularization.
We have used alternative regularizations,
with the four-dimensional
$\gamma_5,~ \gamma_5 =
i \gamma^0 \gamma^1 \gamma^2 \gamma^3$.
After the contributions of $U,~D,~N$ and $E$
are summed up,
the final result of $f^Z_5$
turns out to be unique.

We have checked by an explicit calculation
that in the limit of large fermion masses
all form factors except the four vanish, as
they should, in particular
$\bar{f}^\gamma_1=0$
and $f^\gamma_5 = (-p^2)h^\gamma$.
The results on the remaining four
parameters are listed below:
$$
\eqalign{
\bar{f}^Z_1 = & -{ \alpha \over {32 \pi s^2 c^2}}
 \sum_{i=Q,L} N_i \big\{
{{m^4_{U _i}+ m^4_{D _i}- 6m^2_{U_i} m^2_{D_i}} \over {(m^2_{U_i} -
m^2_{D_i})^2} } + { { 2m^2_{U_i} m^2_{D_i} (m^2_{U_i} + m^2_{D_i})}
\over { (m^2_{U_i} - m^2_{D_i})^3}}~
\ln {{m^2_{U_i}} \over {m^2_{D_i}}} \big\}~,  \cr
\noalign{\vskip4mm}
\bar{f}^\gamma_3
= & - {\alpha \over{24 \pi s^2}}
\sum_{i=Q,L} N_i \big\{
{{m^4_{U_i} +m^4_{D_i} - 14m^2_{U_i}m^2_{D_i}}
\over {2(m^2_{U_i} - m^2_{D_i})^2}} +
{ {3m^2_{U_i}m^2_{D_i} (m^2_{U_i} + m^2_{D_i})}
\over {(m^2_{U_i} - m^2_{D_i})^3}}
\ln {{m^2_{U_i}} \over {m^2_{D_i}}} \big\} \cr
\noalign{\vskip4mm}
& - {\alpha \over{8 \pi s^2}}
\sum_{i=Q,L} N_i Y_i \big \{
{{m^2_{U_i} + m^2_{D_i}}
\over {m^2_{U_i} - m^2_{D_i}}}  -
{ {2m^2_{U_i} m^2_{D_i}} \over {(m^2_{U_i} - m^2_{D_i})^2}}
\ln{{m^2_{U_i}} \over {m^2_{D_i}}}  \big\} ~~, \cr
\noalign{\vskip4mm}
\bar{f}^Z_3
= & -{ \alpha \over{96 \pi s^2c^2}}
 \sum_{i=Q,L} N_i \big\{
{ {5m^4_{U_i} + 5m^4_{D_i} - 46m^2_{U_i}m^2_{D_i}}
\over { (m^2_{U_i} - m^2_{D_i})^2}}  \cr
\noalign{\vskip4mm}
& {\hskip 3.5cm} +{ {18m^2_{U_i}m^2_{D_i} (m^2_{U_i}+m^2_{D_i})}
\over { (m^2_{U_i} - m^2_{D_i})^3}}
\ln{{m^2_{U_i}} \over {m^2_{D_i}}}  \big\}
 -({{s^2} \over {c^2}})\bar{f}^{\gamma}_3 ~~,\cr
\noalign{\vskip4mm}
f^Z_5 =
& -{\alpha \over{32 \pi s^2c^2}}
 \sum_{i=Q,L} N_i \big\{
{ {m^2_{U_i} + m^2_{D_i}} \over {m^2_{U_i} - m^2_{D_i}} } -
{ {2m^2_{U_i}m^2_{D_i}} \over {(m^2_{U_i} - m^2_{D_i})^2}}
\ln{{m^2_{U_i}} \over {m^2_{D_i}}}  \big\} ~~, \cr
}
\eqno{(16)}
$$
where $\alpha \equiv {e^2/{(4 \pi)} }$ and
$ 1-c^2= s^2 \equiv {\sin}^2 \theta_W$, and
$m_{U_Q}=m_U$ and $m_{U_L}=m_N$, etc..

As an illustrative example we
consider an extreme case,
$$
m^2_U \gg m^2_D~~,~~~ m^2_N \gg m^2_E~~.
\eqno{(17)}
$$
It is interesting to note that the
logarithmic dependence disappears,
i.e.,
$$
\eqalign{
\bar{f}^Z_1 & \simeq -
{ \alpha \over {32 \pi s^2 c^2}} (3+1)~~, \cr
\bar{f}^\gamma_3 & \simeq -
{ \alpha \over {24 \pi s^2} } (3-1)~~, \cr
\bar{f}^Z_3 & \simeq -
{\alpha \over {96 \pi s^2c^2}} [ 3 \times (5-4s^2) + (5+4s^2)]~~, \cr
f^Z_5 & \simeq -{ \alpha \over {32 \pi s^2c^2}} (3+1)~~, \cr
}
\eqno{(18)}
$$
where the first term in the parentheses is
the $U, ~D$ contribution, while the second
that of $N,~E$.

(b) {\it Technihadrons}

We have made a rough estimate of the
contributions of technihadrons to the four
parameters by taking the free technifermion picture
and therefore applying eq.(16).  The technifermions
are assumed to have large constituent masses
of order $\Lambda_{TC}$.
Their masses are taken to be degenerate,
$m_{T_U} = m_{T_D}$ and
$m_{T_N} = m_{T_E}$,
in accord with the custodial symmetry
in the technicolor condensation,
$< \bar{T}_U T_U> = < \bar{T}_D T_D>$,
$<\bar{T}_N T_N> = <\bar{T}_E T_E>$.
This picture is suggested by analogy to the result in
QCD that the local average of the $R$-ratio
due to hadrons is well described
by free quark contributions.
For the one generation technicolor model
(with $N_{TC}$ technicolors) we
have obtained
$$
\eqalign{
& \bar{f}^Z_1 = - {\alpha \over {6 \pi s^2 c^2} }N_{TC}~~, \cr
& \bar{f}^\gamma_3 = - {\alpha \over {6 \pi s^2}}N_{TC}~~, \cr
& \bar{f}^Z_3 = - {\alpha \over {6 \pi s^2c^2}}(2-s^2)N_{TC}~~, \cr
& f^Z_5 = 0~~, \cr
}
\eqno{(21)}
$$
where the $f^Z_5$,
being antisymmetric under the exchange of
$m_U \leftrightarrow m_D,~~m_N \leftrightarrow m_E$,
vanishes under the exact custodial symmetry.

To get a rough idea of the
magnitude of the effects of heavy particles in
these four parameters, we take the model of one heavy
fermion generation as an example.  For large
${m^2_U}/{m^2_D} $
and ${m^2_N}/{m^2_E}$
we have from eq.(18) that
$\bar{f}^Z_1 = - {\alpha /{(8\pi s^2c^2)}}
\simeq - 0.002$, which is comparable to
the values of $S$ parameter,
$\alpha S = 2 \alpha / (3\pi) \simeq 0.002$
(for $m_U = m_D$ and $m_N = m_E$).
The experimental determination of $S$
is reaching the precision of this level.
As for the four parameters, it seems difficult to detect
the effects of this minute size in the
first generation experiments of
$e^+e^- \rightarrow W^+W^-$.

\vskip11mm

\centerline{\fourteenbf 4. Summary and remarks}

Any beyond-the-standard model
(new physics) contains heavy particles of
characteristic mass $M$ much larger than
$M_W$.  In some of such models heavy particles
manifest themselves in light particle processes
through radiative
corrections;
their effects do not decouple in the limit
of large $M^2/M^2_W$.
We have studied the non-decoupling effects
in the triple gauge boson (TGB) vertices
$\gamma W^+W^-$ and $Z^0W^+W^-$.
We have sorted out four parameters,
$ \bar{f}^Z_1,~\bar{f}^\gamma_3,~\bar{f}^Z_3,~
f^Z_5$,
which describe the non-decoupling
effects, in comparison with three parameters,
$(S, T, U)$, in the gauge boson two-point functions.

One may wonder whether there exists some
relation between the non-decoupling effects
of gauge boson two-point functions
and those of three-point
functions.
As for the heavy particle contributions of
type i) (see the beginning of this paper),
the coefficient of the
higher dimensional $(d > 4)$ operators is
inversely proportional to the
$SU(2)_L \times U(1)$
singlet large mass $M_S$.
Hence $d=6$ operators are the
most important ones and the two- and three-point
functions are expected to be mutually related [9].

In the type ii) case of heavy particle effects,
we claim that the four parameters
$\bar{f}^Z_1,~\bar{f}^\gamma_3,~\bar{f}^Z_3,~f^Z_5$
and the three parameters $S,~T,~U$ are
independent.  This can be demonstrated
by noting that there exists a set of seven
independent (non-decoupled)
operators consisting of
$D_\mu\phi$ and
${\bf W}_{\mu \nu} \equiv {1 \over 2}\sigma^aW^a_{\mu \nu}$ :
$$
\eqalign{
& (\phi^{\dagger}D_\mu\phi)^2~,~~~(\phi^{\dagger}
{\bf W}_{\mu \nu}\phi)B^{\mu \nu}~,~~~
i~(D_\mu \phi)^{\dagger}(D_\nu \phi)B^{\mu \nu}~~, \cr
& i~(D_\mu \phi)^{\dagger} {\bf W}^{\mu \nu}(D_\nu \phi)~,~~~
(\phi^{\dagger}{\bf W}_{\mu \nu}\phi)^2~,~~~
i~(\phi^{\dagger}{\bf W}^{\mu \nu}\phi)~(D_\mu \phi)^{\dagger}~(D_\nu
\phi)~, \cr
& \epsilon^{\mu \nu \rho \sigma}
{}~(\phi^{\dagger}D_\mu \phi)
{}~[\phi^{\dagger}{\bf W}_{\rho \sigma}D_\nu \phi -
(D_\nu \phi)^{\dagger}{\bf W}_{\rho \sigma}\phi]~~,\cr
}
\eqno{(22)}
$$
modulo
equations of motion.
The seven parameters $\bar{f}^Z_1,~ {\bar{f}}^\gamma_3~, \cdots , U$
are expressed in terms of (linear combinations of) the
coefficients of the seven operators
(22) appearing in the effective Lagrangian.

In our computation of the TGB
couplings we have made no assumption
as to the Higgs mass $m_H$,
and hence our result holds
true for arbitrary values
of $m_H$.
If $m_H$ is light,
the physical Higgs field
$\phi$ appears
in the non-decoupled operators
listed in (22).
If one restricts one's consideration to
the extreme
limit of $m_H \rightarrow \infty$,
one can develop a different
formalism eliminating $\phi$.
In the case
of custodial isospin symmetry and parity
conservation (the
technicolor model is an
example), one can use
the gauged
non-linear $\sigma$-model to examine the problem
we studied above [13].
The task is to find all possible
operators of
$d \leq 4$ in terms of the dimensionless field
$U = \exp (iG^a \sigma^a/2v)$ ($G^a$ are the would-be
Goldstone modes)
and the covariant derivative
$D_\mu$.
The argument has been extended by incorporating
isospin violation effects.
Seven operators are
identified in this approach by Appelquist and Wu [14].

Since $\bar{f}^Z_1, \bar{f}^\gamma_3,
\bar{f}^Z_3, f^Z_5$ are
all dimensionless functions they do not blow up even when
some of heavy particle masses become very large,
in clear contrast to the case
of the $T$ parameter.
The magnitudes
of the four parameters are comparable to that
of $\alpha S$.
Since the four parameters are
independent of the oblique correction
parameters, testing beyond-the-standard
models by measuring these four parameters
in $e^+ e^- \rightarrow W^+ W^-$
experiments is important.
In practice, precise measurements of them in
the coming first generation experiment do not
appear promising.

\vskip8mm

\centerline{\fourteenbf Acknowledgement}
\vskip2mm
We would like to thank Bob Holdom for a critical
reading of the preliminary version
of this paper and helpful discussions on
the approach of the non-linear $\sigma$-model.
We also thank Kenichi Hikasa
and Kiwoon Choi for valuable conversations.
This work is partially supported by Grant-in-Aid
of Ministry of Education, Science and Culture.

\endpage

\noindent
{\twelvebf References}
\item{1.}T. Appelquist and J. Carazzone,
{\sl Phys. Rev.}~{\bf D11}, 2856 (1975).
\item{2.}D.C. Kennedy and B.W. Lynn,
{\sl Nucl. Phys.}~{\bf B322}, 1 (1989).
\item{3.}M.E. Peskin and T. Takeuchi,
{\sl Phys. Rev. Lett.}~{\bf 65}, 964 (1990);
{\sl Phys. Rev.}~{\bf D46}, 381 (1992).
\item{4.}W.J. Marciano and J.L. Rosner,
{\sl Phys. Rev. Lett.}~{\bf 65}, 2963 (1990);
D.C. Kennedy and P. Langacker, {\sl ibid.}
{}~{\bf 65}, 2967 (1990).
\item{5.}G. Altarelli and R. Barbieri,
{\sl Phys. Lett.}~{\bf B253}, 161 (1991).
\item{6.}B. Holdom and J. Terning,
{\sl Phys. Lett.}~{\bf B247}, 88 (1990);
M Golden and L. Randall,
{\sl Nucl. Phys.}~{\bf B361}, 3 (1991);
H. Georgi,
{\sl Nucl. Phys.}~{\bf B363}, 301 (1991).
\item{7.}T. Inami, T. Kawakami and C.S. Lim,
{\sl Mod. Phys. Lett.~{\bf A10}}, 1471 (1995).
\item{8.}B. Grinstein and M. Wise,
{\sl Phys. Lett.}~{\bf B265}, 326 (1991).
\item{9.}A. De R\'ujula, M.B. Gavela,
P. Hernandez and E. Mass\'o,
{\sl Nucl. Phys.}~{\bf B384}, 3 (1992).
\item{10.}C. Ahn, M.E. Peskin, B.W. Lynn
and S. Selipsky,
{\sl Nucl. Phys.}~{\bf B309}, 221 (1988).
\item{11.}K. Hagiwara, K. Hikasa, R. Peccei
and D. Zeppenfeld,
{\sl Nucl. Phys.}~{\bf B282}, 253 (1987).
\item{12.}M. Lemoine and M. Veltman,
{\sl Nucl. Phys.}~{\bf B164}, 445 (1980).
\item{13.}B. Holdom,
{\sl Phys. Lett.}~{\bf B258}, 156 (1991).
C.P. Burgess and D. London,
\tolerance=6000
{\sl Phys. Rev. Lett.}~{\bf 69}, 3428 (1992).
F. Falk, M. Luke and E.H. Simmons,
{\sl Nucl. Phys.}~{\bf B365},
523 (1991).  J. Bagger, S. Dowson and G. Valencia,
{\sl Nucl. Phys.}~{\bf B399}, 364 (1993).
\item{14.}T. Appelquist and G. -H. Wu,
{\sl Phys. Rev.}~{\bf D48}, 3235 (1993).

\endpage
\noindent
{\twelvebf Figure Captions}
\item{Fig. 1}
The $t$-channel ($\nu$-exchange)
and the $s$-channel ($V$-exchange with $V = \gamma$
or $Z^0$) graphs contributing to the
process $e^+e^- \rightarrow W^+W^-$.
The $s$-channel graph has three types
of corrections due to heavy particle loops,
the oblique correction to the $V$ propagator, the
$VW^+W^-$ vertex correction and the
$W^{\pm}$ leg correction, as indicated
by blobs in the figure.

\bye